\documentclass[twocolumn]{aa}
\usepackage{graphicx}
\usepackage{longtable}
\usepackage[authoryear]{natbib}

\usepackage{txfonts}
\def\etal{\hbox{et al.}~}

\def\kms{\,km\,s$^{-1}$}

\def\hal{H$\alpha$}

\def\rsun{{R$_{\odot}$}}

\def\vsini{$v\sin i $}

\begin{document}
   \title{Accretion dynamics and disk evolution in NGC 2264: a study
     based on the Corot\thanks{The CoRoT  space mission was developed
       and is operated by the French space agency CNES, with
       participation of ESA's RSSD and Science Programmes, Austria,
       Belgium, Brazil, Germany, and Spain} photometric observations} 

   \author{S.H.P. Alencar\inst{1},
           P.S. Teixeira\inst{2},
           M.M. Guimar\~aes\inst{1,3},
           P.T. McGinnis\inst{1},
           J.F. Gameiro\inst{4}, 
           J. Bouvier\inst{5},
           S. Aigrain\inst{6,7},
           E. Flaccomio\inst{8}
          \and
           F. Favata\inst{9}
          }

   \offprints{silvia@fisica.ufmg.br}

   \institute{Departamento de F\'{\i}sica -- ICEx -- UFMG, Av. Ant\^onio Carlos, 6627, 30270-901,
              Belo Horizonte, MG, Brazil
         \and ESO, Karl-Schwarzschild-Strasse 2, D-85748 Garching bei M\"unchen, Germany
         \and UFSJ -- Campus Alto Paraopeba -- Rodovia MG 443, KM 7, 36420-000, Ouro Branco, MG, Brazil
	 \and Centro de Astrofisica da Universidade do Porto, Rua das Estrelas, 4150 Porto, Portugal
	 \and Laboratoire d'Astrophysique, Observatoire de Grenoble, BP 53, F-38041 Grenoble C\'edex 9, France
         \and School of Physics, University of Exeter, Exeter, EX4 4QL, UK
         \and Astrophysics, University of Oxford, Denys Wilkinson Building, Oxford, OX4 1DQ, UK
         \and INAF - Osservatorio Astronomico di Palermo, Piazza del Parlamento 1, 90134 Palermo, Italy
	 \and European Space Agency, 8-10 rue Mario Nikis, 75015 Paris, France}

   \date{Received ; accepted }


\abstract
{The young cluster NGC 2264 was observed with the Corot satellite
for 23 days uninterruptedly in March 2008 with unprecedent photometric
accuracy. We present here the first results of the analysis of the 
accreting population that belongs to the cluster and was observed by Corot. 
}
{We intended to look for possible light curve variability of the same
  nature as  
that observed in the classical T Tauri star AA Tau, which was attributed
to a magnetically controlled inner disk warp.
The inner warp dynamics is directly associated with the interaction between
the stellar magnetic field and the inner disk region.
}
{We analysed the Corot light curves of 83 previously known classical T
  Tauri stars that belong to 
NGC 2264 and classified them according to their morphology.
We also studied the Corot light curve morphology as a function of
a \emph{Spitzer}-based classification of the star-disk systems.
}
{The classification derived on the basis of the Corot light curve
  morphology agrees
very well with the \emph{Spitzer} IRAC-based classification of the systems.
The percentage of AA Tau-like light curves decreases as the inner disk dissipates, 
from 40\% $\pm$ 10\% in systems with thick inner disks to 36\% $\pm$ 16\% in systems with anemic disks 
and none in naked photosphere systems. Indeed, 91\% $\pm$ 29\% of the CTTS 
with naked photospheres exhibit pure spot-like variability, 
while only 18\% $\pm$ 7\% of the thick disk systems do so, presumably those seen at low inclination
and thus free of variable obscuration.
}
{
AA Tau-like light curves are found to be fairly common, with a frequency of at least 
$\sim$ 30 to 40\% in young stars with inner dusty disks. The temporal 
evolution of the light curves indicates that the structure of  the inner 
disk warp, located close to the corotation radius and responsible for 
the obscuration episodes, varies over a timescale of a few ($\sim$1-3) 
rotational periods. This probably reflects the highly dynamical nature 
of the star-disk magnetospheric interaction.
}


\titlerunning{Accretion dynamics and disk evolution in NGC 2264}
\authorrunning{Alencar \etal }
\maketitle

\section{Introduction}\label{introduction}
\hspace{1.5em}

T Tauri stars are young, optically visible, low-mass stars still
contracting toward the main sequence. They
have strong magnetic fields ($\sim$2 kG) and are X-ray emitters.
The so-called weak line T Tauri stars (WTTSs) do not exhibit
evidence of disk accretion, while the classical T Tauri
stars (CTTSs) do.
CTTSs present broad emission lines and sometimes also forbidden
emission lines. They are spectroscopically and
photometrically variable, and show ultraviolet (UV), optical and
infrared (IR) excess with respect to the photospheric flux. 

Magnetospheric accretion models are the current consensus to
explain the main observed characteristics of CTTSs
\citep{shu94,har94,muz01,kur06}. In these models, the stellar
magnetosphere is strong enough to disrupt the circumstellar disk
at a few stellar radii. Material in the inner disk, ionized by
stellar radiation, falls towards the star following magnetic
field lines and hits the stellar photosphere at near free-fall velocities,
creating hot spots on the stellar surface. Part of the ionized material
in the inner disk region is ejected in a magnetically
controlled wind. In this scenario, the
broad permitted emission lines are formed partly in the accretion funnel and 
the hot spot emits a continuum flux that is responsible for the
UV and optical excess that veils the photospheric lines.
The IR excess comes from reprocessing by the dust in the disk of the stellar
and accretion radiation and, at least for high accretion rate systems, viscous
heating may also contribute to it. 
When observed, forbidden emission lines are thought to be formed in the 
low density wind.

In the last decade, numerical simulations of accreting young stars
have predicted a very dynamical star-disk interaction, mediated by
the stellar magnetic field \citep{goo99,mat02,rom09,zan09}. 
Many magneto-hydrodynamical (MHD) model
predictions derive from the idea that the stellar magnetic field
interacts with the inner disk region near the co-rotation radius
but not only at the co-rotation point.
Consequently, due to differential rotation between the
star and the inner disk region, the magnetic field lines become distorted
after a few rotational periods and eventually reconnect, restoring the 
initial field configuration. This process goes on as the star rotates.
The time scale for the reconnection events depends on the diffusivity
of the stellar magnetic field lines in the inner disk region, which
is a very poorly constrained parameter.

AA Tau is one of a few CTTSs studied with enough detail 
to test the MHD model dynamical predictions. The star was observed photometrically
for a month during three different campaigns, two of which included high-resolution
simultaneous spectroscopy \citep{bou95,bou03,bou07}. AA Tau shows a light curve (LC) with a flat
maximum interrupted by deep quasi-periodical minima that vary in depth and
width from one rotational cycle to the other. The minima occur with
little color change and are thought to be due to obscuration of the
stellar photosphere  
by circumstellar disk material present in an inner disk warp. The warp is due 
to the interaction of the stellar magnetic field, inclined with respect to the rotation axis, 
and the inner disk region. The spectroscopic results showed that the magnetic field lines 
inflate in the course of a few rotational cycles, as measured by the radial velocity of the red and 
blue absorption components of the \hal \ line. Moreover, when the field lines
are the most inflated, accretion to the star is suppressed, with no apparent
veiling, and low emission-line equivalent widths, which again corroborates the
MHD model predictions. After reconnection, accretion starts all over again.
The last two AA Tau observational campaigns were separated by 5 years and the
field line inflation discovered in the second campaign was still present 
with the same characteristics in the third one.

Although the AA Tau observations gave strong support to the MHD
results, it was still unclear whether this behaviour was common among
CTTSs. In order to test this, one needs good photometric measurements
of a large number of CTTSs, covering several 
rotational cycles. In Taurus the typical rotational period of a CTTS is around 8 days,
so that stars need to be monitored for at least a month continuoulsy.
This is not an easy task from the ground, due to telescope time allocation and 
weather conditions. The Corot satellite additional program to observe the young 
stellar cluster NGC 2264 has allowed us to perform such study on a
large sample of CTTSs.

NGC 2264 is a well studied young stellar cluster with an age of $\approx 3$ Myr located at a 
distance of about 760 pc \citep[see][for a recent review on the cluster]{dah08}. 
We matched the various available datasets on the cluster from the literature based on spatial coincidence
using a 2'' tolerance. In a small number of cases, double identifications 
(i.e. more than one object within 2'') were transformed into single ones,
specifically in cases in which one of the two objects had a small offset and the
other a large one.
Due to the many studies on the cluster, going from UV to X-rays, it was  
possible to establish reasonable criteria for cluster membership.
We considered as likely members of NGC 2264 stars selected according to one
or more of the following criteria: i) photometric \hal \ and variability with 
the data of \citet{lam04} and following their criteria, ii) X-ray detection 
\citep{ram04, fla06} and location on the cluster sequence 
in the ($I$, $R-I$) diagram if $R$ and $I$ magnitudes are available, iii) spectroscopic \hal \ 
equivalent width greater than 10 \AA \ and iv) \hal \ emission line width at
10\% intensity greater than 270 \kms, as proposed by \citet{whi03} to identify
accreting T Tauri stars, and measured
by \citet{fur06} for many cluster members.

The mean rotational period of CTTSs in NGC 2264 is around
3 to 4 days (see Sect. \ref{discussion}). Corot observed the cluster for 
23 days uninterruptedly, covering several rotational cycles for most of the stars,
thus allowing the identification of AA Tau-like candidates and the determination
of precise rotational periods for cluster members. The complete
rotation analysis will be
discussed separately in another paper (Affer et al. in preparation).

\section{Observations}
\hspace{1.5em}

Corot observed NGC 2264 from the 7th to the 30th of March 2008.
The whole cluster fitted in one of the two CCDs normally used for the
exo-planet observations, and stars were observed down to 
$R\simeq 18$. We used the light curves delivered by the Corot pipeline after nominal
corrections \citep{sam07}. We further corrected the
pipeline light curves removing outliers, mainly due to the South Atlantic Anomaly
crossings, by applying a sigma clipping filter, 
taking care not to remove flaring events. The data was also corrected for
the effects due to the entrance and exit into Earth eclipses. We did
not make use of the color information provided by Corot for the
brightest stars, and based our analysis on the broadband, ``white
light'' light curves. All the light curves presented here 
were rebinned to 512 s and correspond to the integrated flux in the Corot mask.
The flux RMS over 512 s achieved is of the order of 0.0005 for a $R=12$ star and
0.004 for a $R=16$ star, yielding the most detailed light curves of young accreting systems 
up to now, as shown in Figure \ref{LC_days}. A complete catalog of the observations will be
published in a subsequent paper (Favata et al. in preparation).


\section{Results}
\hspace{1.5em}

\subsection{Classical T Tauri sample selection}\label{selection}

\begin{figure*}[htb]
\centering
\includegraphics[width=18cm]{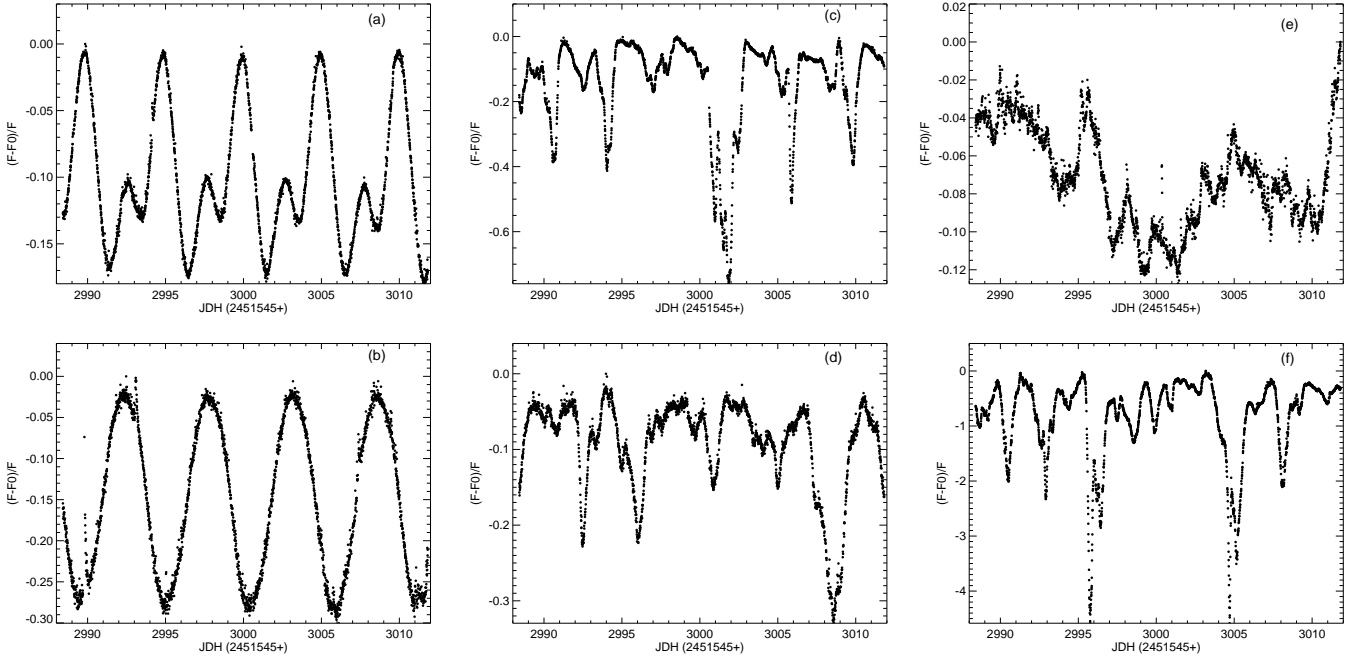}
\caption{A sample of 6 CTTS light curves from the COROT observation of
  NGC 2264. $F_0$ corresponds to the maximum flux value
of each LC. Light curves $a$ and $b$ have been classified as spot-like,
$c$ and $d$ as AA Tau-like and $e$ and $f$ as irregular.
}
\label{LC_days}
\end{figure*}

After selecting the cluster members as described in Sect. \ref{introduction}, 
we classified as CTTSs stars that presented
either \hal\ equivalent width 
greater than 10 \AA, $U-V$ excess less than a threshold calculated below, or \hal\ width at
10\% intensity greater than 270 \kms. Some stars presented more than
one of the above characteristics. 
\citet{whi03} showed that the accretion criterium based on \hal \ equivalent
width is spectral-type dependent, the threshold being smaller than 10 \AA \
for spectral types earlier than K7 and greater than 10 \AA \ for spectral types 
later than M2.5. Whenever spectral type information was available, we 
followed the accretion criteria proposed by \citet{whi03}, instead of using the
standard 10 \AA \ value.
\citet{reb02} proposed that stars in NGC 2264 with $U-V$ excess less than $-0.5$ could
be classified as disk candidates. This threshold value was based on a study of the 
Orion star forming region \citep{reb00}. However, the $U-V$ excess is also expected to be 
spectral-type dependent, stars with later spectral types presenting higher $U-V$ excess
compared to earlier ones, for the same mass accretion rate, due to the higher contrast
between the hot spot and the stellar photosphere in later spectral types.
Therefore we looked for all the stars of our sample that were selected as CTTS based on 
$U-V$ excess and another criterion (\hal\ equivalent width or \hal\ width at 10\%
intensity). We separated them in two spectral type ranges (K0-K6 with 14 stars and K7-M3 
with 10 stars) and computed the mean $U-V$ excesses in each dataset. The K0-K6 CTTSs
present $U-V$ excess of $-1.06 \pm 0.48$ mag and the K7-M3 CTTSs present $U-V$ excess of 
$-1.69 \pm 0.57$ mag. Using the one sigma upper boundary in each spectral type
range as a threshold to separate CTTS from WTTS ($-0.58$ for K0-K6 and $-1.12$ for K7-M3), 
we select 7 stars as CTTSs based only on $U-V$ excess. While the value of $-0.5$ mag
proposed by \citet{reb02} seems to be adequate for stars in the K0-K6 spectral range, it is
apparently too high to select K7-M3 CTTSs in NGC 2264.

We found 83 CTTSs among the 301 observed cluster members and present in 
Table \ref{classification} the data used to make the CTTS classification.
The \hal \ equivalent width was taken from \citet{reb02} and \citet{dah05} except for
six stars (CID 223957455, 223959618, 223964667, 223968688, 223991832, 223994721) 
for which we measured the \hal \ equivalent width ourselves, using the high 
resolution hectoechelle spectra kindly provided by Gabor F{\H u}r{\'e}sz. 
The $U-V$ excess data was obtained from \citet{reb02} and
\citet{fal06} (the data table was kindly provided by Cassandra Fallscheer)
and the \hal \ width at 10\% intensity comes from \citet{fur06}.

Some CTTSs that were selected based on their $U-V$ excess have \hal \ values that
are below 10 \AA \ and would therefore not be selected as CTTS based only on
\hal \ equivalent width. However, their $U-V$ excess is lower than our established 
threshold values and in the same range 
as many other systems that present either \hal \ equivalent width greater
than 10 \AA \ or \hal \ width at 10\% larger than 270 \kms. We have to be aware 
that both \hal \ equivalent width and $U-V$ excess are strongly variable in these stars and
were not measured simultaneously. So it seems reasonable to use both criteria
to select possible CTTSs.

\subsection{Morphological light curve classification}\label{morphology}
We looked for periodical variations in the LCs of the observed CTTSs, using
the Scargle periodogram as modified by \citet{hor86}, and found that 51 out of
83 CTTSs presented periodical variability.
Periodic LCs were divided in two groups:
group PI, containing sinusoidal-like LCs with stable shape from cycle to
cycle, and 
group PII, flat-topped LCs with a clear maximum interrupted
by minima that can vary in width and depth from cycle to cycle.
The variations in group PI are associated with 
long-lived spots with lifetime of at least of weeks, while group PII
is associated with AA Tau-like systems, where most of the variability is due to
obscuration by circumstellar disk material.
The non-periodical LCs (group NP) can be due to obscuration by non-uniformly
distributed circumstellar material or to non-steady accretion or both.

\begin{figure}[htb]
\centering
\includegraphics[width=9cm]{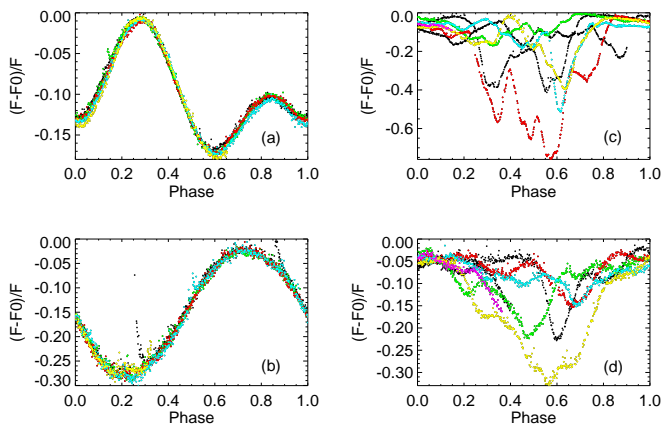}
\caption{The four periodic LCs from Fig. \ref{LC_days} folded in phase. Different colors
correspond to different cycles.
}
\label{LC_phase}
\end{figure}

A total of 83 CTTSs that belong to NGC 2264 were observed by Corot, 28 of
which were classified as spot-like (group PI), 23 as AA Tau-like (group PII) 
and 32 as irregular (group NP).
A sample of light curves of each type is shown in Fig. \ref{LC_days}. In Figure \ref{LC_phase} we
present the periodical LCs of Fig. \ref{LC_days} folded in phase with the
periods determined with the Scargle periodogram as modified by \citet{hor86}. We can notice the stability
of the spot-like LCs (Fig. \ref{LC_days} and Fig. \ref{LC_phase} $a$ and $b$) in the timescale 
of the observations, which makes them, in general, easily distinguishable from the 
AA Tau-like ones (Fig. \ref{LC_days} and Fig. \ref{LC_phase} $c$ and $d$). 
Among the irregular LCs, some look more like due to variable accretion events (variable hot spots, 
Fig. \ref{LC_days} $e$) and others to obscuration by non-uniform circumstellar material (Fig. \ref{LC_days} $f$), 
but it is hard to decide which process is the dominant one based only on the Corot light curves
without any color information. Therefore we did not make any attempt
to further classify the irregular systems.

We measured the variability amplitude of a LC as $(({\rm Flux_{max}} - {\rm Flux_{min}})/{\rm Flux_{median}}) \times 100$.
In our sample, the observed CTTS variability amplitudes range from 3\% to 137\%, excluding
flaring events.
The variability amplitude of spot-like LCs is generally around 10\% to 15\% and
most of the stars that present more than 20\% of variability amplitude have LCs that
are classified as due to obscuration by circumstellar material (AA Tau-like). 
The AA Tau system variability amplitude, measured from data in the literature, is 76\% 
and 8 out of 83 stars observed by Corot have a higher variability amplitude than AA Tau. 
This shows that, although AA Tau presents a high variability amplitude, probably
due to its high inclination with respect to our line of sight \citep[$\sim$ 75\degr][]{bou03}, 
it is not exceptional among CTTS systems.

From the results in this Section, it becomes clear that the AA Tau photometric behavior is 
not an exception, but a rather common occurence among young stellar systems, 
representing 28\% $\pm$ 6\% of the CTTS systems in NGC 2264 observed with Corot. 
The percentage of AA Tau-like systems among the observed CTTSs seems reasonable, given that only 
some geometric configurations (i.e. high inclination) would produce occultation of
the stellar photosphere, and the chances for occultation events will also depend 
on the disk warp's location and scale-height. 
This result is however higher than the value of
10\% to 15\% calculated by \citet{ber00} using flared disk models for the propability
of observing partial occultation events in CTTS systems. However, the disk models by 
\citet{ber00} did not include an inner disk warp, and consequently underestimate the 
probability of obscuration.
Assuming a random distribution of axial inclinations, the fraction of AA Tau-like 
LCs in our sample ($\geq 28\%$) suggests $h/r \sim 0.3$ for inner disk warps, where $h$ 
and $r$ are the inner warp height and distance to the central star. This is larger than 
the standard value used in $\alpha$-disk models, where $h/r \sim 0.05-0.1$ \citep{ber88,duc10}.


\subsection{Merging Corot and \emph{Spitzer} data}\label{corot_spitzer}
\emph{Spitzer} IRAC data were also available for the cluster \citep{tei08},
with a total of 68 CTTSs present in both \emph{Spitzer} and Corot observations.
IRAC is useful to identify near-infrared excess emission that arises from warm dusty 
circumstellar material. 
We used the $\alpha_{\rm IRAC}$ index, which represents the slope of the
spectral energy distribution between 3.6 $\mu$m and 8 $\mu$m, to classify
the inner disk structure of the observed systems, following the criteria proposed
by \citet{lad06}. Stars with $\alpha_{\rm IRAC} < -2.56$ are
classified as naked photospheres (i.e., these systems are devoid of dust within a few AU), 
stars with $-2.56 < \alpha_{\rm IRAC} < -1.80$
have anemic disks (optically thin inner disks), stars with
$-1.80 < \alpha_{\rm IRAC} < -0.5$ have optically thick inner disks (referred to as 
thick disks henceforth), those with $-0.5 < \alpha_{\rm IRAC} < 0.5$ are flat spectra 
sources and the ones with $\alpha_{\rm IRAC} > 0.5$ are Class I objects.

\begin{figure*}[htb]
\centering
\includegraphics[width=18cm]{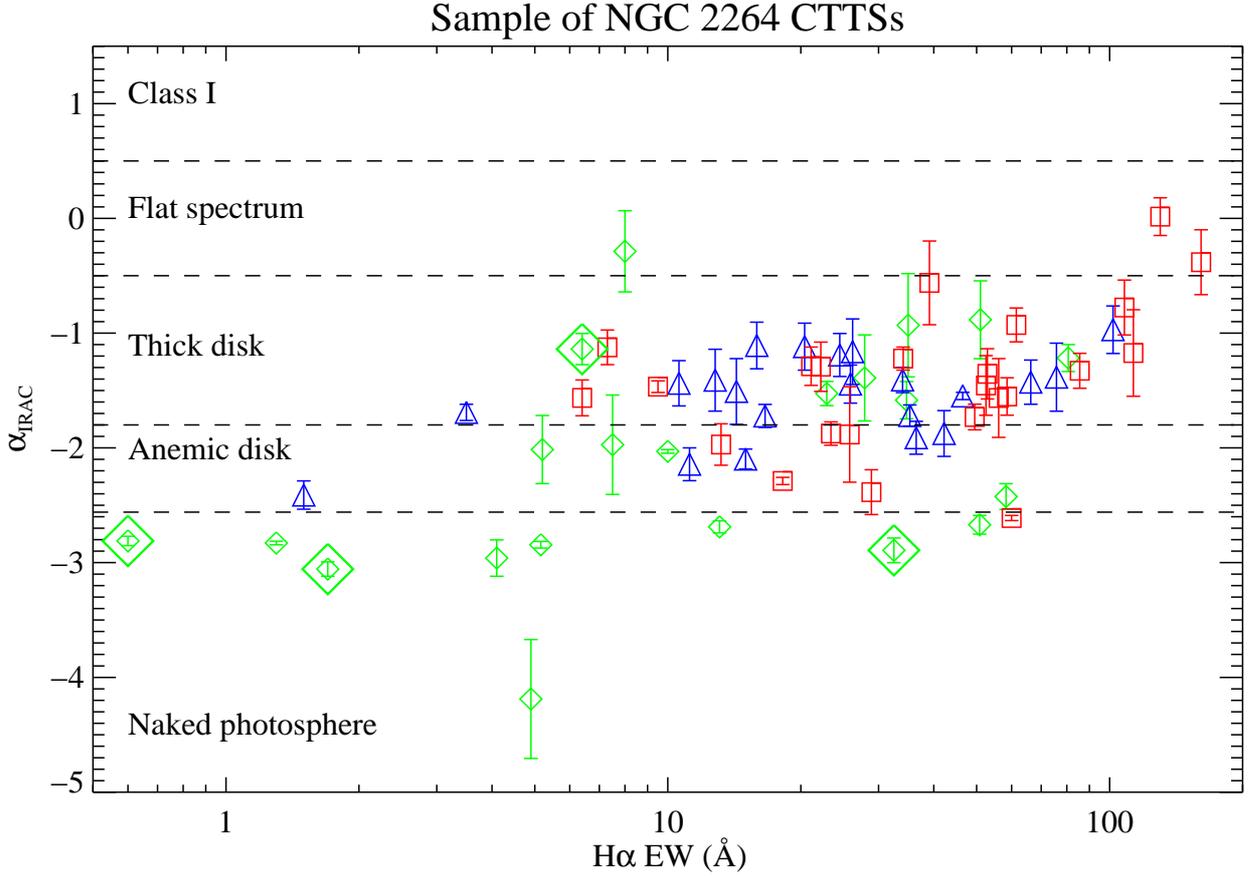}
\caption{NGC 2264 classical T Tauri stars observed with
Corot and \emph{Spitzer}. Diamonds correspond to spot-like LCs, triangles
to AA Tau-like LCs and squares to non-periodical LCs. The four stars with large
overplotted diamonds are fast rotators ($p < 2$ days). One fast rotator was not
included in this figure (CID 500007354, $p= 1.17$ days), because of the
huge uncertainty (2.89) on its $\alpha_{\rm IRAC}$ index. It is nevertheless included
in Table \ref{classification} and in our LC analysis.
}
\label{spitzer}
\end{figure*}

We combined both datasets in order to see if the light curve
morphology was related to the evolution of the inner disk structure.
The result, presented in Figure \ref{spitzer}, shows that the agreement between the two
classification approaches of CTTS (based on the Corot light curves and
on the \emph{Spitzer} photometry) is excellent. None of the systems
with naked photospheres exhibit AA Tau-like  
LCs and 10/11 (91\% $\pm$ 29\%) of them show pure spot-like variability. The percentage of 
AA Tau-like LCs, which are due to obscuration by circumstellar disk material,
thus increases from 0\% for the evolved inner disk systems 
(naked photospheres) to 36\% $\pm$ 16\% for the anemic disk systems and 40\% $\pm$ 10\% for 
the thick disk systems. Because some of the non-periodic LCs may also be 
partly caused by circumstellar obscuration (see a possible example in Fig. \ref{LC_days}$f$), 
the fractional estimates of 
obscuration LCs are lower limits for systems with anemic and thick 
disks.

Spot-like systems represent 91\% $\pm$ 29\% of the naked photospheres, 
28\% $\pm$ 14\% of the anemic disks and 18\% $\pm$ 7\% of thick disk systems.
They are observed in systems where there is no obscuring material in our
line of sight towards the star. This could be due to inner disk clearing, corresponding 
to spot-like LCs which represent naked photosphere systems and some anemic
disks, shown in Fig. \ref{spitzer}.
Spot-like LCs could also come from low inclination systems. In this case, there could 
be inner disk material, but it would not obscure the star. This probably corresponds to some 
anemic disk systems and to all thick disk systems that show spot-like LCs.

The non-periodical LCs represent 9\% $\pm$ 9\% of the naked photospheres, 36\% $\pm$ 16\% of the anemic disks
and 42\% $\pm$ 10\% of the thick disk systems.
The irregular LCs among naked photosphere systems can be due to non-steady accretion, 
which will produce short-lived and variable hot spots. Some irregular systems among anemic 
disks and thick disks are also likely due to a combination of non-steady accretion and low 
inclination, as apparently observed in TW Hya \citep{ruc08}. It may however not be straightforward 
to assign non-periodical LCs to non-steady accretion, as recently shown by \citet{kul09}. They 
computed MHD models of the interaction between a magnetized
star and its circumstellar disk with non-steady accretion and showed that, 
at large misalignment angles ($\theta > 25$\degr) between the stellar rotation and magnetic axis, 
hotspots are approximately fixed on the star's surface, even during strongly unstable accretion, and 
consequently the LCs always show the stellar rotation period. 
Assuming that the circumstellar environment of a CTTS may be complex, random accretion events 
due to circumstellar blobs, which fall towards the star, could also temporarily occult the star 
and explain some of the observed non-periodical systems, as proposed by \citet{dew03} to explain 
the irregular photometric variability of the CTTS SU Aur.
Another situation that could produce irregular LCs is a flared disk seen at high inclination.
We could then expect to see partial obscuration 
by circumstellar disk material from the disk outer layers. \citet{ber00} calculated the 
partial obscuration probability by a flared disk and showed that, for a typical CTTS, it would be 
of the order of 10\% to 15\%. In this case, assuming Keplerian disk rotation and 
a typical CTTS disk with an outer radius of 100 AU, we would unfortunately not be able to measure 
short scale variability, as observed with Corot, due to material located in most disk radii because
of the limited duration of our observations.

\section{Discussion}\label{discussion}
\begin{figure}[htb]
\centering
\includegraphics[width=9cm]{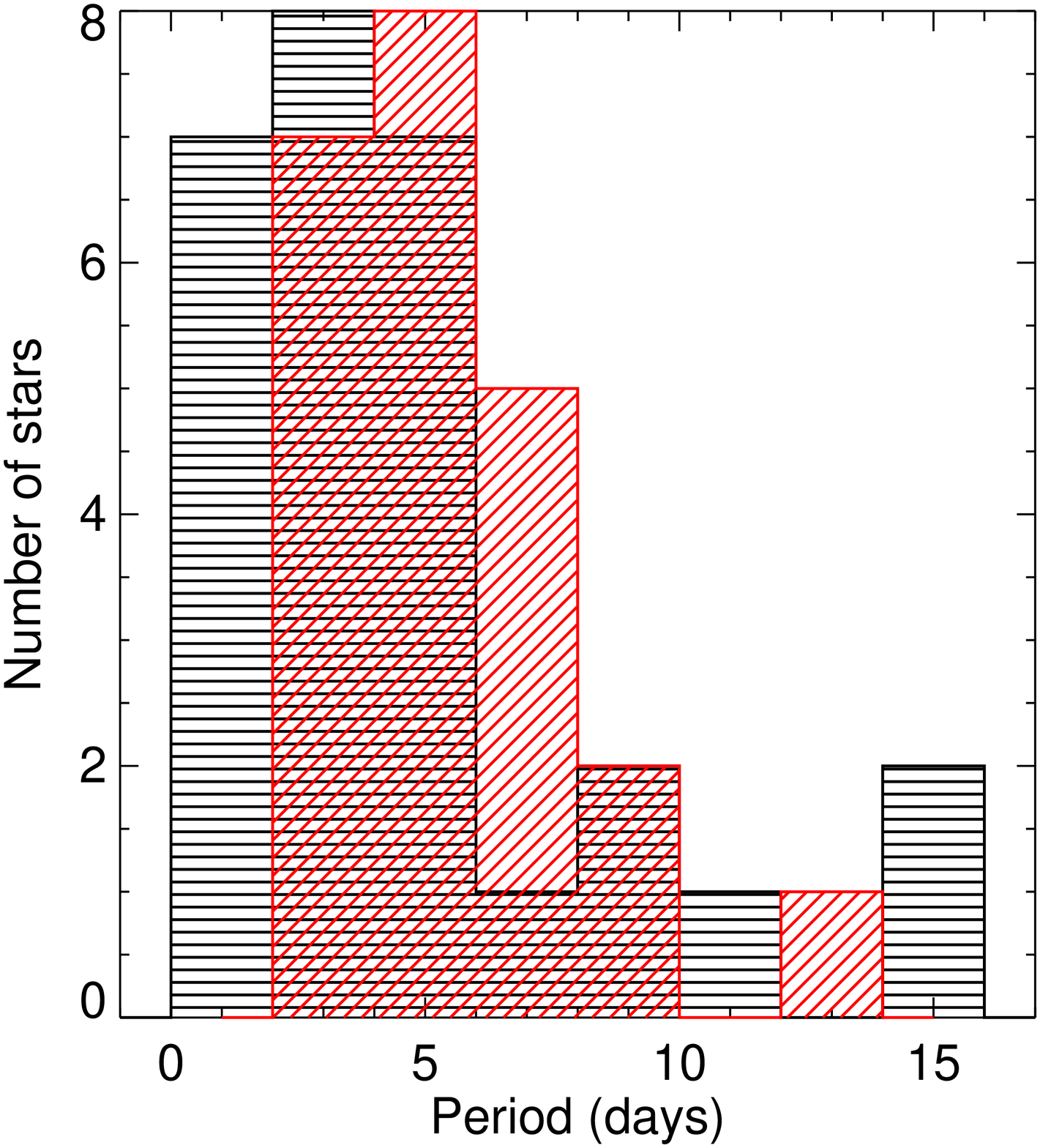}
\caption{Period distribution of CTTSs. The black histogram corresponds to spot-like LCs and
the red histogram to the AA Tau-like LCs.
} 
\label{periods}
\end{figure}

We calculated periods for all CTTSs that presented periodic variations
(51 out of 83 CTTSs), using the Scargle periodogram as modified
by \citet{hor86}. 
We present in Fig. \ref{periods} the period distribution of the 
spot-like (black histogram) and AA Tau-like (red histogram) systems.
A complete discussion on period distribution in NGC 2264 is
presented in another paper (Affer et al. in preparation).

The major difference between the period distributions
of spot-like and AA Tau-like systems is that fast rotators
($p < 2$ days) are only found with spot-like LCs. 
Among the seven fast rotating CTTSs, five also have \emph{Spitzer} IRAC measurements. 
Four of them are classified as naked photosphere systems
and another is a thick disk system.
The thick disk system (CID 223980447, $\alpha_{\rm IRAC} = -1.14 \pm 0.13$, $p=1.67$ days) 
is a K6 star \citep{dah05} and has high-resolution hectoechelle spectra \citep{fur06} 
that Gabor F{\H u}r{\'e}sz kindly made available to us.  We measured \vsini$=30$ \kms, 
using the SYNTH3 code provided by Dr. Oleg Kochukhov \citep{koc07}, together with MARCS 
atmospheres \citep{gus08} and atomic lines from VALD \citep{kup00,kup99}. Assuming $R_*=2$ \rsun, 
we obtain $i=29.4$\degr. This low inclination can explain the presence of circumstellar material
with no obscuration in the LC.
Except for this system, most fast rotators are found among systems that have 
cleared out their inner disk regions, which could be an indication that 
as the star-disk coupling decreases, stars tend to spin-up, as also found by \citet{reb06}
and \citet{cie07} in Orion and NGC 2264. This is not
a straghtforward conclusion, however, from our data, since the number of fast rotators
is small and on the other hand some naked photosphere and anemic disk 
systems are actually found to rotate slowly, with periods up to 10 to 15 days. 
The periods measured in the AA Tau-like LCs fall in the range of 
periods obtained from spot-like LCs.
The periods measured from the spot-like LCs correspond to stellar rotational
periods, since spots are located at the stellar photosphere. This indicates
that AA Tau-like periods are within the range of stellar rotational periods of CTTSs in NGC 2264 
and therefore the material that obscures the star must be located close to the corotation radius.
Since the inner warp is by definition located close or at the dust disk truncation radius, 
this implies that the dust truncation radius is near the corotation radius in the systems
we classified as AA Tau-like. 
\citet{car07} showed that the inner gas radius is on average slightly smaller than the
corotation radius, while the inner dust radius falls at or outside the corotation radius.
This is quite consistent with the Corot results.

Since the inner disk warp is located near the corotation radius, the variations 
observed from cycle to cycle in width and depth of the photometric minima should 
be related to the dynamical star-disk interaction in the inner disk region, that 
is thought to be responsible for the accumulation of material near the disk truncation 
region, forming inner disk warps. Like AA Tau,
the star-disk interaction is seen to be very dynamic on a rotational
timescale, as predicted by MHD models of young magnetized star-disk systems
\citep{goo99,mat02,rom09,zan09}. 
In our observations some systems are more regular and stable than
others, but it is quite common to see systems that present photometric
minima that vary substantially from cycle to cycle, still keeping their
overall periodic nature (see Figs.\ref{LC_days},\ref{LC_phase} $c$ and $d$).

\section{Conclusions}
We showed that the AA Tau photometric behavior is common among CTTSs,
being present in 28\% $\pm$ 6\% of the CTTSs in our sample. 
This represents a lower limit, since the AA Tau-like LCs are more likely 
produced at high inclinations, and we are probably missing about 20\% to 30\% 
of the very high inclination systems, according to  the calculations
by \citet{ber00}, which 
will be totally obscured by a flared disk and thus too faint to be observed by Corot.

If our interpretation of such systems is correct, the photometric minima are due to
obscuring material located in the inner disk region, near the corotation radius.
This material is built up through the interaction between an inclined stellar magnetosphere and 
the inner disk region. The observed periodical changes in width and depth of the 
observed minima, over a timescale of a few ($\sim$ 1-3) rotational periods, 
would then reflect the dynamics of such an interaction,
as predicted by MHD models of young low mass star-disk systems.

We compared the Corot light curves with \emph{Spitzer} IRAC data of the same
systems and showed that the agreement between classifications based on
the two datasets is excellent.
The percentage of AA Tau-like light curves, which are due to obscuration
by circumstellar material in the inner disk region, varies as the inner disk dissipates,
decreasing from 40\%  $\pm$ 10\% in systems with thick inner disks to 
36\% $\pm$ 16\% in systems with anemic disks
and none in naked photosphere systems. Indeed, 91\% $\pm$ 29\% of the systems
with naked photospheres exhibit pure spot-like variability,
while only 18\% $\pm$ 7\% of the thick disk systems do so, presumably those seen at low 
inclination and thus free of variable obscuration.

\hspace{1.5em}


\begin{acknowledgements}
We thank Gabor F{\H u}r{\'e}sz for making his Hectoechelle spectra and the
updated electronic version of his tables available to us.
We thank Cassandra Fallscheer for making available to us her table
of U-V excess measurements.
This research is based on data collected on the Corot satellite.
SHPA and MMG acknowledge support from CNPq, CAPES and Fapemig.
JFG acknowleges support from the FCT project PTDC/CTE-AST/66181/2006.
\end{acknowledgements}


\clearpage \onecolumn
\begin{longtable}{llllllllllll}
\caption{\label{classification} Classical T Tauri stars that belong to NGC 2264 and were observed by Corot.
The \hal \ equivalent widths come from \citet{reb02} and \citet{dah05}, except for
six stars (CID 223957455, 223959618, 223964667, 223968688, 223991832, 223994721)
for which we measured the \hal \ equivalent width ourselves, using the high
resolution hectoechelle spectra kindly provided by Gabor F{\H u}r{\'e}sz. The $U-V$ excesses
were taken from \citet{reb02} (UV1) and \citet{fal06} (UV2, the U-V excess data table was provided by Cassandra
Fallscheer). Stars that present
\hal \ width at 10\% intensity larger than $270$ \kms \ are identified by a x sign
in column eight, following the classification of \citet{fur06}. The LC group
and period values were determined in the present work. The $\alpha_{\rm IRAC}$ index
is described in Sect. \ref{corot_spitzer} and comes from \citet{tei08}.}\\
\hline\hline
RA        & Dec      & V     & Corot ID  & \hal EW & UV1    & UV2    & \hal10\% & SpT & LC group & Period & $\alpha_{\rm IRAC}$\\
(deg)     & (deg)    & (mag) &           & (\AA)   & (mag)  & (mag)  &          &     &          & (days) &       \\ \hline
\endfirsthead
\caption{continued.}\\
\hline\hline
RA        & Dec      & V     & Corot ID  & \hal EW & UV1    & UV2    & \hal10\% & SpT & LC group & Period & $\alpha_{\rm IRAC}$\\
(deg)     & (deg)    & (mag) &           & (\AA)   & (mag)  & (mag)  &          &     &          & (days) &       \\ \hline
\endhead
\hline
\endfoot
 99.89338 &  9.91424 & 16.64 & 223957455 & 22.1    & ...    &  ...   & x        & ... & NP       & ...    & ...   \\
 99.92281 &  9.77214 & 14.64 & 223959618 & 28.4    & ...    &  ...   & x        & ... & PII      & 3.87   & ...   \\
 99.99684 &  9.45681 & 16.34 & 223964667 & 42.2    & ...    &  ...   & x        & ... & PII      & 6.45   & -1.87 \\
100.04666 &  9.63501 & 16.67 & 223968039 & 52.9    & -1.15  &  ...   & x        & K6  & NP       & ...    & -1.35 \\
100.05673 &  9.41375 & 13.02 & 223968688 & 0.60    & ...    &  ...   & x        & K0  & PI       & 1.11   & -2.81 \\ 
100.05710 &  9.94183 & 17.02 & 400007614 & 130.2   & ...    &  ...   & x        & M2  & NP       & ...    & 0.0149\\
100.09620 &  9.46176 & 16.04 & 223971231 & 49.5    & -1.62  &  ...   & x        & K5  & NP       & ...    & -1.73 \\        
100.09889 &  9.92330 & 17.55 & 223971383 & 80.6    & ...    &  ...   & x        & ... & PI       & 4.66   & -1.22 \\
100.12006 &  9.51718 & 15.20 & 223972652 & 39.1    & -0.89  &  ...   & ...      & K4  & NP       & ...    & -0.56 \\
100.12186 &  9.73542 & 17.69 & 400007809 & 31.3    & -1.36  &  ...   & x        & M2  & NP       & ...    & ...   \\
100.12859 &  9.57794 & 16.03 & 223973200 & 22.2    & ...    &  ...   & x        & K1  & NP       & ...    & -1.29 \\        
100.13013 &  9.51864 & 15.43 & 223973292 &  1.7    & -0.91  &  ...   & ...      & K6  & PI       & 1.96   & ...   \\         
100.15217 &  9.84601 & 16.67 & 400007538 & 21.1    & -2.11  & -1.687 & ...      & M2  & NP       & ...    & -1.29 \\ 
100.15262 &  9.80638 & 15.62 & 500007298 &  4.9    & -1.43  & -0.606 & ...      & K5  & PI       & 15.70  & -4.19 \\
100.15781 &  9.58167 & 16.64 & 400007528 & 23.4    & ...    &  ...   & x        & M3  & NP       & ...    & -1.87 \\ 
100.16297 &  9.84961 & 15.43 & 500007252 & 46.5    & -1.89  &  ...   & x        & K4  & PII      & 7.06   & -1.55 \\
100.16840 &  9.84735 & 15.50 & 500007272 & 58.3    & ...    &  ...   & ...      & K7  & PI       & 3.76   & -2.42 \\
100.17086 &  9.46509 & 16.53 & 500007505 & 13.2    & ...    &  ...   & ...      & ... & NP       & ...    & -1.97 \\
100.17095 &  9.79936 & 16.01 & 500007379 &  7.5    & -1.22  & -0.984 & ...      & K7  & PI       & 14.15  & -1.97 \\
100.17216 &  9.85066 & 15.70 & 500007315 & 24.5    & -2.13  &  ...   & ...      & K7  & PII      & 7.80   & -1.19 \\
100.17233 &  9.90385 & 13.57 & 223975844 & 12.2    &  0.02  &  ...   & x        & G3  & PI       & 3.31   & ...   \\        
100.17435 &  9.86237 & 15.49 & 500007269 & 23.4    & ...    &  ...   & ...      & K5  & PI       & 3.75   & -1.53 \\
100.17576 &  9.56040 & 12.09 & 223976028 &  7.3    & ...    &  ...   & x        & G0  & NP       & ...    & -1.12 \\
100.18006 &  9.78535 & 16.33 & 500007460 & 27.1    & ...    &  ...   & x        & K6  & NP       & ...    & ...   \\
100.18580 &  9.54061 & 18.13 & 500007930 & 60.0    & ...    &  ...   & ...      & M3  & NP       & ...    & -2.61 \\
100.18819 &  9.47901 & 14.55 & 223976747 &  7.2    & -0.16  &  ...   & x        & K2  & PII      & 3.16   & ...   \\
100.19793 &  9.82471 & 14.14 & 500007120 & 12.8    & ...    & -0.806 & x        & K1  & PII      & 4.23   & -1.41 \\
100.19968 &  9.55087 & 18.24 & 500007963 & ...     & ...    & -0.917 & ...      & ... & PI       & 2.56   & ...   \\  
100.20505 &  9.96077 & 17.46 & 500007730 & 50.8    & ...    &  ...   & ...      & M1  & PI       & 11.85  & -2.67 \\
100.20789 &  9.61375 & 15.37 & 223977953 & 66.3    & -1.79  & -1.906 & ...      & K4  & PII      & 4.96   & -1.43 \\        
100.21081 &  9.91593 & 15.13 & 500007209 & 11.2    & ...    &  ...   & x        & K2  & PII      & 2.51   & -2.14 \\
100.21326 &  9.74615 & 12.05 & 223978308 &  3.5    & ...    &  ...   & x        & G0  & PII      & 5.40   & -1.69 \\
100.22346 &  9.55686 & 14.70 & 223978921 & 18.2    & ...    & -0.643 & x        & K1  & NP       & ...    & -2.29 \\
100.22607 &  9.82232 & 16.39 & 500007473 & 161.1   & -2.11  &  ...   & x        & M0  & NP       & ...    & -0.38 \\
100.22990 &  9.84718 & 15.89 & 500007354 &  2.8    & -0.64  & -0.774 & ...      & K5.5 & PI      & 1.17   & -9.15 \\
100.23215 &  9.85385 & 15.54 & 500007283 & 8.00    & ...    &  ...   & ...      & K5.5 & PI      & 3.23   & -0.29 \\
100.23663 &  9.63029 & 15.09 & 223979728 & 113.2   & ...    & -1.353 & x        & M1  & NP       & ...    & -1.17 \\
100.24208 &  9.61483 & 18.05 & 223980048 & 34.0    & ...    &  ...   & ...      & ... & PII      & 4.06   & -1.41 \\
100.24447 &  9.60368 & 17.40 & 223980233 & 22.2    & ...    &  ...   & ...      & M4  & NP       & ...    & ...   \\
100.24510 &  9.65522 & 17.02 & 223980258 &  27.9   & ...    &  ...   & x        & M0  & PI       & 7.05   & -1.39 \\
100.24516 &  9.51592 & 14.04 & 223980264 & 14.3    & ...    &  ...   & x        & K2.5 & PII      & 3.46   & -1.51 \\
100.24770 &  9.99596 & 15.34 & 223980412 & 7.41    & -0.53  &  ...   & ...      & K5  & PI       & 3.23   & ...   \\
100.24792 &  9.49770 & 17.00 & 500007610 & 26.2    & ...    &  ...   & ...      & M3  & PII      & 4.66   & -1.16 \\
100.24811 &  9.58636 & 15.58 & 223980447 &  6.4    & -1.01  & -0.427 & x        & K6  & PI       & 1.67   & -1.14 \\
100.25209 &  9.75088 & 15.39 & 223980688 & 15.0    & -0.59  & -0.723 & x        & K3  & PII      & 4.16   & -2.10 \\
100.25214 &  9.48776 & 14.56 & 223980693 & 16.6    & ...    &  ...   & x        & K4  & PII      & 5.35   & -1.72 \\
100.25323 &  9.85620 & 14.63 & 500007157 &  1.6    & ...    & -0.658 & ...      & K1  & PI       & 4.36   & ...   \\
100.25408 &  9.54568 & 13.51 & 223980807 &  6.4    & ...    &  ...   & x        & K1  & NP       & ...    & -1.56 \\  
100.25767 &  9.64475 & 14.70 & 223981023 &  1.5    & ...    &  ...   & x        & K4  & PII      & 7.05   & -2.41 \\
100.26266 &  9.62660 & 19.17 & 500008211 & 34.1    & ...    &  ...   & ...      & M1  & NP       & ...    & -1.22 \\
100.26503 &  9.50806 & 17.67 & 400007803 & 20.4    & ...    &  ...   & ...      & ... & PII      & 9.75   & -1.12 \\
100.26789 &  9.41449 & 15.79 & 500007335 & 101.8   & -2.34  &  ...   & x        & M0  & PII      & 7.36   & -0.97 \\
100.26905 &  9.64190 & 17.88 & 500007857 & 108.    & ...    &  ...   & ...      & M3  & NP       & ...    & -0.78 \\
100.27071 &  9.84613 & 14.36 & 223981811 & 36.5    & -0.29  & -0.816 & x        & K1  & PII      & 3.73   & -1.91 \\
100.27124 &  9.86239 & 15.39 & 500007248 &  1.7    & -0.75  & -0.570 & ...      & K5  & PI       & 1.88   & -3.06 \\
100.27583 &  9.60638 & 13.52 & 223982136 & 10.0    & ...    &  ...   & x        & G3  & PI       & 3.01   & -2.03 \\
100.27595 &  9.41769 & 18.01 & 500007896 & 34.7    & ...    &  ...   & ...      & M5  & PI       & 9.30   & -1.58 \\
100.27679 &  9.47745 & 17.30 & 400007686 &  56.1   & ...    &  ...   & x        & M1.5 & NP      & ...    & -1.57 \\
100.27808 &  9.57943 & 15.97 & 500007369 & 49.4    & ...    &  ...   & x        & G6  & NP       & ...    & ...   \\      
100.28734 &  9.56278 & 17.53 & 500007752 &  51.0   & ...    &  ...   & x        & M3  & PI       & 4.01   & -0.88 \\
100.29582 &  9.59881 & 17.44 & 500007727 & 61.5    & ...    & -2.037 & x        & K7  & NP       & ...    & -0.93 \\
100.30241 &  9.87533 & 14.07 & 500007115 & 35.3    & ...    &  ...   & x        & G   & PII      & 2.01   & -1.72 \\
100.30362 &  9.43746 & 13.76 & 500007089 & 85.6    & ...    &  ...   & x        & K4  & NP       & ...    & -1.33 \\
100.31035 &  9.62065 & 17.23 & 500007667 &  4.1    & ...    & -0.697 & ...      & ... & PI       & 5.41   & -2.96 \\
100.32188 &  9.90899 & 15.64 & 223985009 & 58.3    & -2.40  &  ...   & ...      & K7  & NP       & ...    & ...   \\
100.32467 &  9.48364 & 18.55 & 500008049 & 231.4   & ...    &  ...   & ...      & M2.5 & NP      & ...    & -0.68 \\
100.32534 &  9.64038 & 18.58 & 500008061 & 32.5    & ...    &  ...   & ...      & M3  & PI       & 0.98   & -2.89 \\
100.32613 &  9.56488 & 15.02 & 223985261 & 28.9    & -0.97  &  ...   & x        & K4  & NP       & ...    & -2.39 \\
100.33752 &  9.56005 & 15.20 & 223985987 & 10.6    & ...    &  ...   & x        & K6  & PII      & 3.31   & -1.44 \\
100.34851 &  9.78788 & 17.93 & 500007872 &  5.2    & ...    & -0.767 & ...      & ... & PI       & 8.20   & -2.01 \\
100.35227 &  9.62653 & 17.38 & 400007709 &  8.9    & -0.99  &  ...   & ...      & M3  & PI       & 0.76   & ...   \\
100.35677 &  9.57861 & 16.15 & 223987178 & 15.9    & -0.61  & -1.233 & ...      & M0  & PII      & 4.96   & -1.11 \\
100.36250 &  9.50365 & 17.47 & 400007734 & 25.8    & ...    &  ...   & ...      & M1  & NP       & ...    & -1.88 \\
100.37968 &  9.44951 & 14.19 & 500007122 & 25.9    & ...    &  ...   & x        & ... & PII      & 12.53  & -1.44 \\
100.38169 &  9.80912 & 14.63 & 223988742 & 5.16    & -0.08  & -0.409 & ...      & K2  & PI       & 5.03   & -2.84 \\
100.38331 &  10.0068 & 15.59 & 223988827 & 13.1    & ...    &  ...   & ...      & K5  & PI       & 4.78   & -2.69 \\
100.38543 &  9.63540 & 14.66 & 223988965 &  1.3    & -0.67  &  ...   & ...      & K6  & PI       & 3.23   & -2.83 \\
100.39397 &  9.60904 & 17.16 & 223989567 &  4.5    & -0.83  &  ...   & x        & M1  & NP       & ...    & ...   \\
100.40536 &  9.75186 & 15.44 & 223990299 & 35.0    & ...    & -0.826 & x        & K4  & PI       & 4.51   & -0.93 \\
100.41155 &  9.53661 & 15.40 & 500007249 & 58.6    & ...    &  ...   & x        & K4  & NP       & ...    & -1.55 \\ 
100.41564 &  9.67443 & 13.86 & 223990964 & 52.5    & ...    & -1.061 & ...      & K4  & NP       & ...    & -1.46 \\ 
100.42867 &  9.41900 & 16.65 & 223991832 & 75.8    & ...    &  ...   & x        & ... & PII      & 8.40   & -1.38 \\
100.47104 &  9.96747 & 14.99 & 223994721 &  9.5    & ...    &  ...   & x        & K7  & NP       & ...    & -1.47 \\
\hline
\end{longtable}
\end{document}